\documentclass[preprint2]{aastex}

\newcommand {\ia}{\'\i }

\def\nh{{N$_{\rm H}$}}
\def\xmm{{\em XMM--Newton}}

\def\cm2{{cm$^{-2}$}}

\shorttitle{On the Anticorrelation Between Galaxy Light Concentration and X--ray-to-Optical Flux Ratio} 
\shortauthors{Povi\'c et al.}

\begin{document}

\title{On the Anticorrelation Between Galaxy Light Concentration and X--ray-to-Optical Flux Ratio}
 
\author{M. Povi\'c}
\affil{Instituto de Astrof\'isica de Canarias, 38205  La Laguna,  Spain}
\email{mpovic@iac.es} 
\author{M. S\'anchez-Portal}
\affil{Herschel Science Centre, ESAC/INSA, P.O. Box 78, 28691 Villanueva de la Ca\~nada, Madrid, Spain}
\email{miguel.sanchez@sciops.esa.int}
\author{A. M. P\'erez Garc\'ia\altaffilmark{1}}
\affil{Instituto de Astrof\'isica de Canarias, 38205  La Laguna,  Spain}
\author{A. Bongiovanni\altaffilmark{1}}
\affil{Instituto de Astrof\'isica de Canarias, 38205  La Laguna,  Spain}
\author{J. Cepa\altaffilmark{2}}
\affil{Departamento de Astrof{\ia}sica, Universidad de La Laguna, 38205  La Laguna,  Spain}
\author{M. Fern\'andez Lorenzo}
\affil{Instituto de Astrof\'isica de Canarias, 38205  La Laguna,  Spain}
\author{M. A. Lara-L\'opez}
\affil{Instituto de Astrof\'isica de Canarias, 38205  La Laguna,  Spain}
\author{J. I. Gonz\'alez-Serrano}
\affil{Instituto de F{\ia}sica de Cantabria, CSIC-Universidad de Cantabria, Santander, Spain}
\author{E. J. Alfaro}
\affil{Instituto de Astrof{\ia}sica de Andaluc{\ia}a-CSIC, Granada, Spain}


\altaffiltext{1}{Departamento de Astrof{\ia}sica, Universidad de La Laguna, 38205  La Laguna,  Spain}
\altaffiltext{2}{Instituto de Astrof\'isica de Canarias, 38205  La Laguna,  Spain}

\begin{abstract}

Active Galactic Nuclei (AGN) play an important role in many aspects of the modern cosmology, and of particular interest is the issue of the interplay between AGN and their host galaxy. Using X--ray and optical data sets, we have explored the properties of a large sample of AGNs in the Subaru/XMM-Newton Deep Survey (SXDS) field, and studied their evolution in relation with the evolution of their host galaxy. We present here an anticorrelation between X--ray-to-optical flux ratio (X/O) and galaxy light concentration (C), which has been found for the first time and might suggest that early type galaxies, having poor matter 
supply to feed the AGN activity, have lower Eddington rates than those of late type galaxies.

\end{abstract}


\section{Introduction}

Deep X--ray extragalactic surveys constitute an extremely useful tool to detect the population of active galactic nuclei (AGN), to investigate their time evolution, and to shed light on their triggering mechanisms. Moreover, hard X--ray surveys are capable to detect all but the most absorbed (Compton thick) AGNs. Therefore, they provide the most complete and unbiased samples of AGNs. The combination of X--ray data with optical observations allows succesfully tackling the study of the AGN population \citep{geor06,steffen06,barcons07}. Several works \citep{gebhardt00,graham01,kauff03} have shown that the AGNs are directly related with some host galaxy properties, particularly with their bulges. A tight correlation between black hole (BH) mass and the bulge dispersion velocity has been well established \citep{ferrarese00,gebhardt00}, as well as with the galaxy light concentration \citep{graham01}.\\
We present here, for the first time, an anticorrelation between the galaxy light concentration and the X--ray-to-optical flux ratio for a large sample of AGNs in the Subaru/XMM Deep Survey (SXDS)\footnote{http://subarutelescope.org/Science/SubaruProject/SDS/} field, 
showing that more concentrated galaxies (earlier morphological types) have lower X/O ratio. This anticorrelation is found to be significant in a wide range of photometric redshifts, with a similar first grade polynomial function fitting in different redshift bins. We suggest that accretion rates in early type objects (having higher values of concentration index) in tipically gas-poor environment, are lower than in late type objects, that are surrounded by larger reserves of gas for AGN feeding.

\section{Observational data}
\label{observational_data}

\subsection{X--ray data processing and source detection}
\label{xray_data}

The \xmm\  observations of SXDS field \citep{furusawa08} were collected from the \xmm\ Science Archive (XSA). Seven pointings around the central coordinates ($\alpha$\,=\,02$^h$ 18$^m$00$^s$, $\delta$\,=\,-05$^{\circ}$00$'$00$''$; PI. M. Watson) have been gathered, covering a total area of $\sim$\,1.3\,deg$^2$.\\
The EPIC observations were reprocessed using SAS v7.1.2 standard reduction procedures. Six energy ranges were defined: soft (0.5-2 keV), hard-2 (2-4.5 keV), very\_hard  (4.5-10 keV), very\_hard-2 (4-7 keV), total (0.5-10 keV) and total-2 (0.5-7 keV). Hardness ratios were defined as follows:
\begin{equation}
HR\left(\Delta_1E,\Delta_2E\right) = \frac{CR\left(\Delta_1E\right) - CR\left(\Delta_2E\right)}{CR\left(\Delta_1E\right) 
+ CR\left(\Delta_2E\right)}
\end{equation}
\noindent where $\Delta_1E$ y $\Delta_2E$ are different energy bands and $CR\left(\Delta_nE\right)$ is the
count rate in a given energy band. The hardness ratio used in this work is HR$_{h2s}$\,$\equiv$\,HR(hard-2, soft).\\
Source detection was performed by means of the {\tt edetect\_chain} SAS procedure with a likehood parameter 
$ML = -\ln (1-P) > 14$, where $P$ is the probability that the source exists. We considered only those sources for which $S/N > 2$ in the total band. We have detected 1121 sources fulfilling these conditions. Limiting flux is 1.3\,$\times$\,$10^{-15}$\,erg\,cm$^{-2}$s$^{-1}$ in the 0.5--4.5\,keV range.

\subsection{Optical counterparts}
\label{catalog}

Optical data have been gathered from the SXDS Public Suprime-Cam Data Release 1 (DR1)
\citep{furusawa08}. The total mapped area is 1.22 deg$^2$, centered at 
$\alpha$\,=\,02$^h$18$^m$00$^s$, $\delta$\,=\,-\,05$^o$00$'$00$''$. Limiting AB magnitudes are
28.4, 27.8, 27.7, 27.7 and 26.6 in B, V, R$_c$, i$^{\prime}$ and z$^{\prime}$ filters,
respectively. We have characterized the morphology of extended objects by means of 
the Abraham concentration index \citep[C; ][]{abraham94}, computed by a modified version of SExtractor 2.5.0 \citep[][modifications from B. Holwerda]{bertin96}, as the ratio between the integrated flux within a radius defined by the
normalized radius $\alpha$=0.3, and the total flux. \\

We cross-matched the X--ray and optical catalogs with a search radius of 3$''$, obtaining a completeness of 99.9\% and
a reliability of 76.2\% \citep[applying de Ruiter methodology, see ][for more details]{deruiter77,Povic}. We have found 808 X--ray sources with optical counterpart.

\section{Photometric redshifts and K-corrections}
\label{photoz_kcorr}

We have used ZEBRA \citep{Feldmann} to compute photometric redshifts of the optical counterparts. Input photometric data includes 11 optical and IR bands:
Subaru/SuprimeCam B, V, R$_c$, i$^{\prime}$ and z$^{\prime}$; J, H and K from the UKIDSS\footnote{The UKIDSS project is defined in \cite{lawrence07}. UKIDSS uses the UKIRT Wide Field Camera (WFCAM).
The photometric system is described in \cite{hewett06}, and the calibration is described in \cite{hodgkin09}. The pipeline processing and science archive are described in Irwin et al (2009, in prep) and \cite{hambly08}. We have used data from the 3th data release, which is described in detail in \cite{warren08}.} survey; Spitzer/IRAC DR2 data in 3.6$\mu$m 4.5$\mu$m and 5.8$\mu$m bands \citep{surace05}. The best SED template set has been gathered from the SWIRE template library \citep{polletta07}. As a complementary check, photometric redshifts 
have been also computed using the HyperZ code \citep{Bolzonella} with the same template set, accepting only those objects for which the redshift differences from both codes are less than 0.1
for z\,$<$\,1 and 0.2 for z\,$\geq$\,1. From the initial sample of 808 sources with optical counterparts, 377 objects fulfill these conditions.  

In order to obtain K-corrections of the optical fluxes we have used the \textsc{idl} routine \textsc{kcorrect}  \citep{blanton}.
On the other hand, the X--ray fluxes were K-corrected by assuming a standard power law SED with $\Gamma$\,=\,1.8. A rough classification between unabsorbed and absorbed AGNs has been made using HR$_{h2s}$, that is quite sensitive to absorption \citep{della04,dwelly05}; those objects with HR$_{h2s}$\,$\le$\,0.35 
(some 53\%) have been considered as unabsorbed, and no intrinsic absorption has been applied to the power law SED; otherwise the objects have been considered as absorbed and a fixed intrinsic absorption  
\nh\,=\,1.0\,$\times$\,10$^{22}$\,cm$^{-2}$ has been included.

\section{The anticorrelation between galaxy light concentration and X/O flux ratio}
\label{sec_rel}

\begin{figure*}[ht!]
\centering
\includegraphics[angle=0,width=0.85\textwidth]{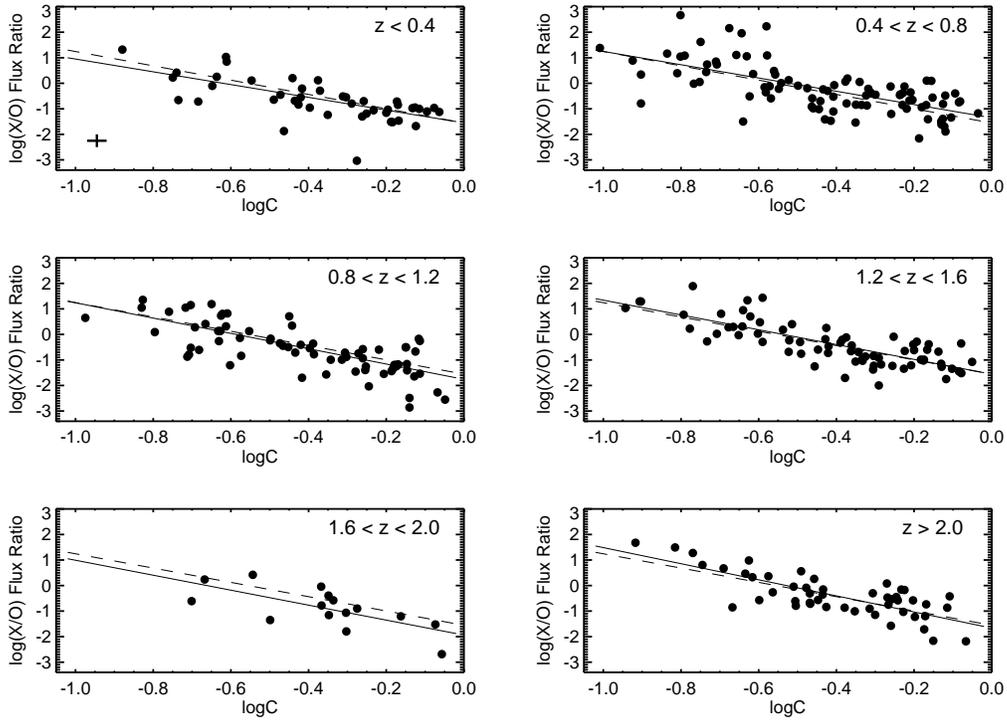}
\protect\caption[ ]{\small{Relationship between the concentration index and the K-corrected X/O ratio in six redshift bins. The continuum line corresponds to the linear fit for data points in each redshift bin, while the dashed line represents the fit for the whole sample. Final error bars are included in the first plot, and are significant in all six redshift ranges. We found that C is accurate to $\sim$5\%. The average error of X/O ratio is found to be $\Delta$X/O\,=\,0.351 for the full sample. Goodness of full sample fit, as measured by the Spearman's rank correlation coefficient ($\rho$) is -0.708.}  \label{XO_C}}
\end{figure*}

Correlating the X--ray and optical properties we found a very strong anticorrelation between the X/O ratio\footnote{computed as $F_{\rm 0.5-4.5keV}/F_R$, where the optical flux $F_R$ has
been derived from the SExtractor ``auto'' magnitudes in the R band} and the concentration index, represented in Fig.~\ref{XO_C} \citep[already 
observed in the population of AGNs of the Groth field, see ][]{Povic, sanchez}. The X/O ratio measures the X--ray flux (in 0.5--4.5\,keV band), normalized with the R-band optical flux of the whole galaxy (nucleus, bulge and disc), although the sources in our sample are predominantly nucleus/bulge dominated. For the whole sample, a first order polynomial function has been fitted with a slope of -2.76\,$\pm$\,0.14 and log X/O-intercept of -1.54\,$\pm$\,0.07. Both unabsorbed and absorbed AGNs follow the same relation. Final C and X/O error bars are shown in the first plot of Fig.~\ref{XO_C}, and are significant in all six redshift ranges. In general, we obtained that C is accurate to 5\%, mainly due to the uncertainty introduced by the nuclear-point source influence (see Section \ref{cubsec_bias}). The average error of X/O flux ratio is $\Delta$X/O\,=\,0.351 for the full sample.

\subsection{Analysis of possible bias effects}
\label{cubsec_bias}

\begin{figure*}[ht!]
\centering
\includegraphics[angle=0,width=0.85\textwidth]{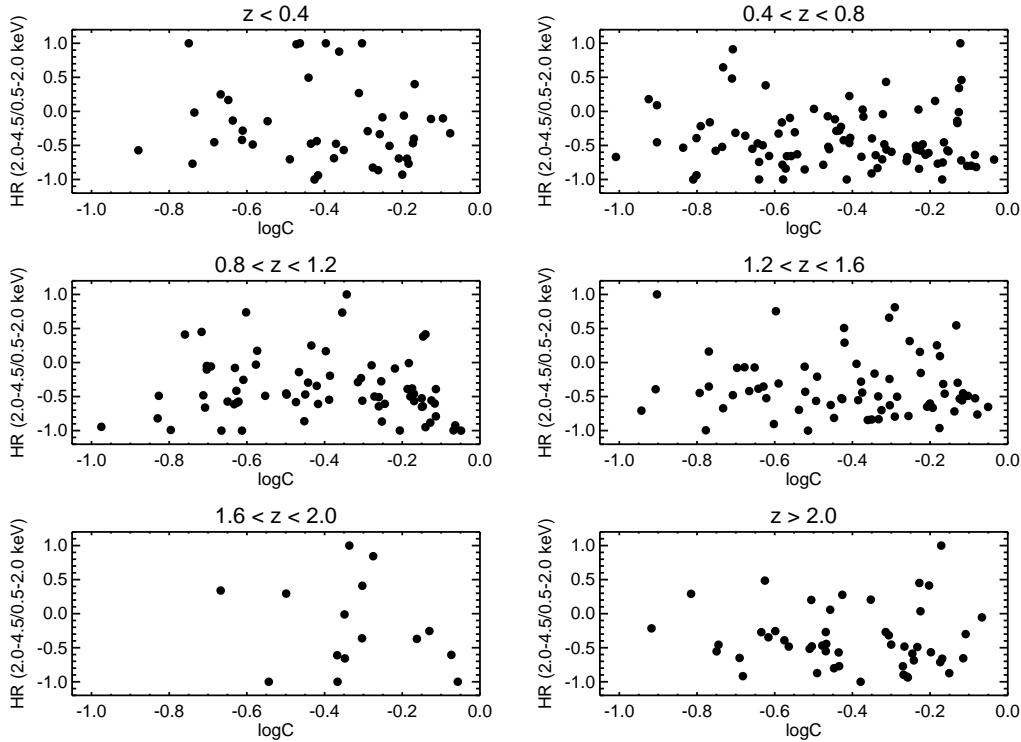}
\protect\caption[ ]{\small{Relationship between the concentration index C and the HR$_{h2s}$ hardness ratio in the same redshift bins as in Fig.~\ref{XO_C}. }  \label{HR_C}}
\end{figure*}

\textit{- Redshift and Bandpass shift.} It is expected that distance will directly affect the measurements of C, tending to shift objects to apparently later Hubble types \citep{Brinchmann98, Conselice03}; besides this, there is an influence of a bandpass shift with respect to the rest wavelength \citep{Bershady00}. We obtained K-corrected X/O ratios and we created six redshift ranges, as shown in Fig.~\ref{XO_C}. The range of C (and X/O) is the same in all six bins, i.e. basically independent of redshift. Moreover, the relation is the same in all redshift ranges, as verified by linear regression fits. These facts 
indicate that the effect of these biases is negligible.\\
\textit{- Obscuration.} Intrinsic absorption of the nuclear point-source could make the X/O flux ratio larger and C smaller, leading to apparent anticorrelation between these two parameters. We used HR$_{h2s}$ criterion in order to examine the absorption effect on our anticorrelation. As already mentioned in Section \ref{photoz_kcorr}, this parameter should select efficiently the majority of the most heavily absorbed sources. 
Fig.~\ref{HR_C} shows the relationship between HR$_{h2s}$ and C in the same six redshift ranges as in Fig.~\ref{XO_C}. No correlation was found in any redshift range. So, even if HR$_{h2s}$ might not be effective to discriminate absorbed AGNs at high redshifts, we can at least claim that for objects within z\,$\lesssim$\,1.5  \citep[e.g. ][]{Akylas06} obscuration should not play an important role in the observed anticorrelation.\\
\textit{- Nuclear point source.} The presence of a nuclear point-like source could affect the determination of C, in the sense that the measured values are overestimated for the host galaxy comparing with their true values. \cite{gabor09} found that the presence of a nuclear point source could have a significant bias on concentration measurements to higher values, although the authors do not quantify it. However, the influence of the nuclear point source in the concentration indexes has been investigated in HST counterparts of Chandra sources by \cite{grogin03}, finding that the AGN host concentration differences are truly related to differences in the host galaxy structure and not to the nuclear point-source flux contribution, quantifying the possible influence at C in $\sim$\,10\%. We have extended this analysis to ground-based data, by building several galaxy models (elliptical, lenticular, early, intermediate and late spirals), convolved with an HST-like PSF (a gaussian function with FWHM\,$=$\,0.043\,arcsec) and a ground-like PSF (FWHM\,$=$\,1.0\,arcsec) representative of our data, adding in both cases different fractions of an unresolved nuclear source (nucleus/total fraction up to 50\%) and deriving the corresponding Abraham concentration indexes. Even though the range of C of ground-like data shrinks considerably with respect to that observed in HST-like data,  we observe that HST and ground concentration indexes are very tightly linearly correlated and, that the variation of C is dominated by the host galaxy morphology and not by the nuclear component contribution. Both results suggest that the influence of this bias in the anticorrelation is small. We derive that the maximum  effect of the nuclear point source on C is about 7\% in HST-like models and less than 3\% in ground-like ones.\\
\textit{- Seeing, Apparent Brightness and Size.} Optically fainter objects tend to show lower concentration indexes, and this effect is enhanced with poorer seeing. In fact, this correlation is observed in our data: the apparent magnitude and logC are linearly correlated\footnote{Actually this correlation is twofold: besides the observational bias discussed here, it is observed that less concentrated galaxies tend to be intrinsically fainter \citep[e.g.][]{blanton01}; this trend is also found in our data.}. A weak tendency that apparently smaller objects have lower concentration indexes is present in our data, affecting mainly late type galaxies. Besides the observational bias effect, this tendency could be also partially intrinsic (as in the case of the apparent brightness), showing that intrinsically smaller objects have lower C. In order to test if C is still a good tracer of the host galaxy morphology, we have performed an independent classification based on visual inspection \citep[see][]{Povic}. There is a good agreement between the results derived from the combination of the concentration and asymmetry indexes and those obtained by inspection. Therefore, even taking into account that apparent brightness and seeing effects play an important role, we are still confident that C is a good tracer of galaxy morphology, and that, on average, lower concentration indexes are tracing later morphological types. Moreover, we performed a preliminary analysis of public HST/ACS images of the EGS field \citep{davis07,lotz08}, obtaining the I-band optical fluxes and concentrations; we combined the optical information  with the AEGIS EGS X--ray public data \citep{laird09}, obtaining a clear anticorrelation between the X/O ratio and C, in all four AEGIS X--ray energy bands.

\subsection{Comparison with galaxy mock \\catalogs}
\label{virgo_millennium}

\begin{figure*}[ht!]
\centering
\includegraphics[angle=0,width=0.89\textwidth]{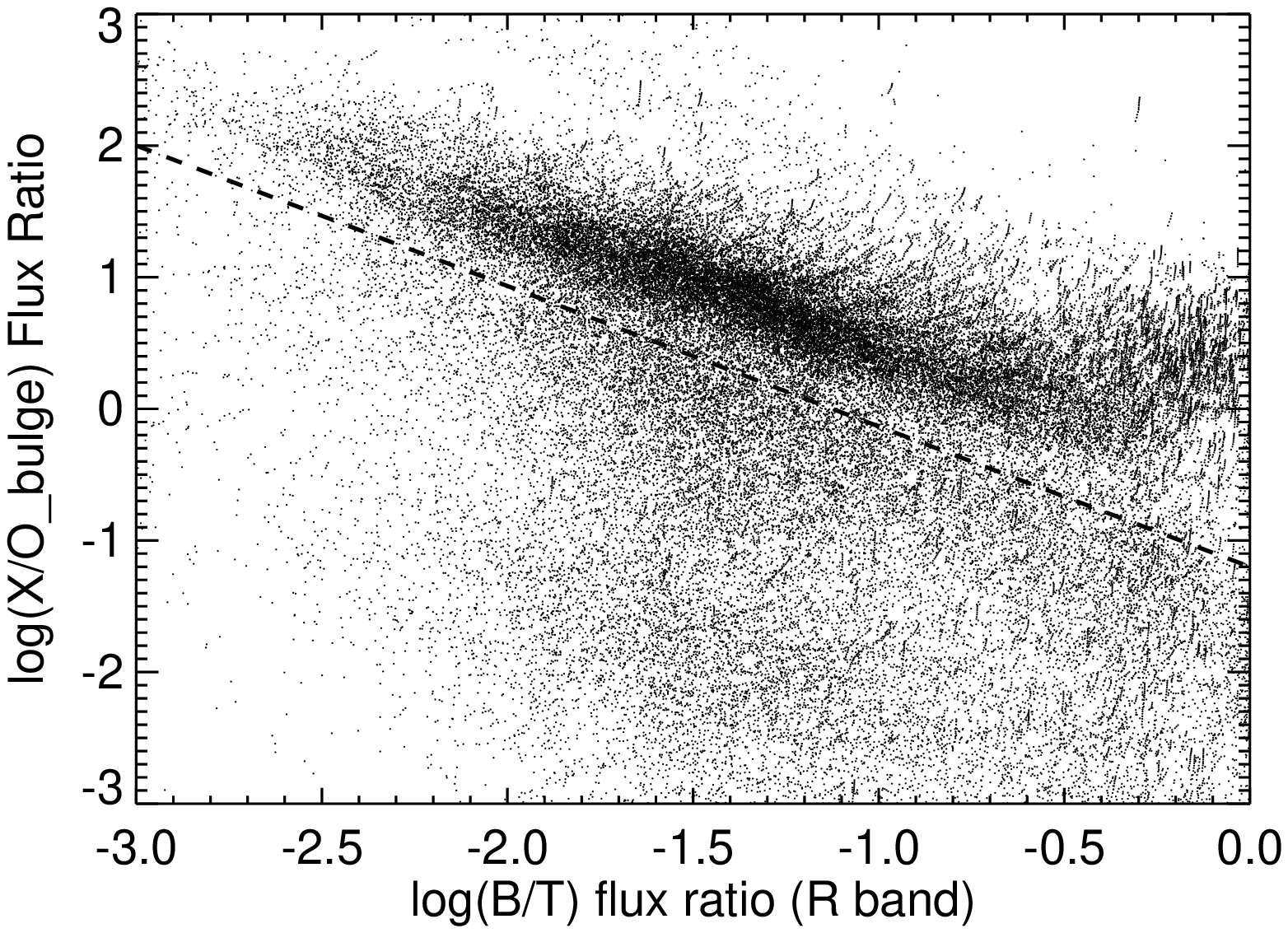}
\protect\caption[ ]{\small{
Relationship between the X--ray-to-bulge optical flux ratio and bulge-to-total flux ratio (R band) for The Millennium-II simulated galaxies \citep{delucia07} in the redshift range z\,$\le$\,0.4. The same relationship is obtained in the other five redshift ranges shown in Fig.~\ref{XO_C}; 99\% of the objects located below the dashed line have X--ray bolometric luminosities less than 10$^{42}$\,erg\,s$^{-1}$, i.e. below the faint end of our sample.
}  \label{fig_virgo_mill}}
\end{figure*}

We used The Millennium-II Simulations database  \citep[][and references therein]{boylan-kolchin09} in order to perform further tests of the existence of an anticorrelation between X/O and C. 
The mock catalog we used, contained in the MPAGalaxies Database, is the table DeLucia2006a \citep{croton06,delucia07}. We worked with more than 700000 galaxies, dividing them in the same redshift ranges as in Fig.~\ref{XO_C}. We obtained the R-band bulge-to-total flux ratio (B/T) for all simulated galaxies; this parameter is proportional to the galaxy light concentration. The X/O ratios are calculated from the catalog's X--ray bolometric luminosities and the R-band bulge fluxes. Fig.~\ref{fig_virgo_mill} represents the obtained relationship for more than 60000 simulated galaxies with z\,$\le$\,0.4. A clear anticorrelation can be seen for all B/T values except for pure bulge galaxies (i.e. with log(B/T)\,$>$\,-0.25), whose real population should be very small, so we do not expect to have a significant number of these galaxies in our real sample. In the log(B/T) interval $\approx$\,[-1.5,-0.2], where most of our objects are located, this clear anticorrelation can be confirmed, showing that less-concentrated or less bulge-dominated objects (later Hubble types) have higher X/O$_{bulge}$ flux ratios. Finally, for galaxies with log(B/T)\,$\lesssim$\,-1.5, we can expect to be below our X--ray detection limits. There is a significant number of outliers, mostly found below the observed anticorrelation. All these objects show properties of small, dwarf or satellite galaxies (small maximum rotational velocitiy,  reduced virial and stellar mass and hot/cold gas content, low SFR) and low bolometric X--ray luminosities, well below 10$^{42}$\,erg\,s$^{-1}$, making most of them undetectable at our sensitivity thresholds. 
In our sample, only 3\% of all X--ray emitters with optical counterparts have X--ray luminosities lower than 10$^{42}$\,erg\,s$^{-1}$, in the observed 0.5-4.5\,keV energy range. Figure~\ref{fig_virgo_mill} shows a clear difference in X--ray luminosity between the galaxies belonging to the anticorrelation sequence, and to the outliers. However, there are examples of outliers found observationally (see Section \ref{sec_discussion}). 

\section{Discussion}
\label{sec_discussion}

All the tests performed, described in Sections \ref{cubsec_bias} and \ref{virgo_millennium}, suggest that the observed anticorrelation cannot be solely attributed to biases, but also to a true physical effect. Deriving a possible explanation requires understanding the physical connections between the observables C and X/O. \\
There is a tight correlation between the  C$_{31}$ concentration index (reflecting morphologies of bulges) and the mass of the central supermassive BH, showing that more concentrated galaxies have more massive BHs \citep{graham01,graham01a}. The Abraham concentration index C used in this work is not the most effective tracer of bulge morphologies;  however, all the tests  performed so far suggest that C is a parameter good enough to trace the galaxy light concentration and that, on average, higher concentration indexes are tracing more concentrated objects (or earlier morphological types), while lower concentration indexes are tracing later morphological types. Moreover, C and C$_{31}$ concentration index are monotonically correlated. We analyzed the morphology of SXDS X--ray emitters with optical counterparts using GALSVM \citep{HuertasCompany08}. We obtained Gini and Moment of light (M$_{20}$), finding a clear correlation between these parameters and C, indicating again that higher values of C are tracing earlier Hubble types. On the other hand, there is a tight correlation between Gini and M$_{20}$ with C$_{31}$ concentration index \citep[e.g. ][]{lotz04}.\\
On the other hand, the physical connections of the X/O flux ratio are not so evident, but as shown below  it could be interpreted as an indicator of the Eddington ratio. As already mentioned, our sample is predominantly made of nucleus/bulge dominated sources. Thus we can distinguish between two possible cases:\\
- If the nuclear luminosity dominates that of the host galaxy, the X/O ratio can be thought as a 
measure of the X--ray to optical spectral index, $\alpha_{OX}$\footnote{This relation is usually defined as $\alpha_{OX} = 0.3838 \log (f_{2keV}/f_{2500\AA})$ }. On the other hand, there is a strong correlation between the hard/soft X--ray spectral index, $\alpha_X$, and the Eddington ratio \citep[e.g. ][]{grupe04,bian05}. The relationship between $\alpha_X$ and $\alpha_{OX}$ has not been thoroughly studied in large samples of AGN, but a linear correlation has been found between both spectral indexes in
a sample comprising 22 out of 23 quasars in the complete the Palomar-Green X--ray sample with z\,$<$\,0.4 and M$_B$\,$<$\,-23 \citep{shang07}. If this last correlation holds for our sample, X/O could be tracing the Eddington ratio in the large nuclear luminosity limit.\\
- If the host galaxy luminosity is large when compared with the nuclear luminosity ($L_{bulge+disc} \gg L_{nucleus}$), the X/O ratio  can be thought as a lower limit of the X--ray-to-bulge luminosity ratio, that is in turn a (weak) measure of the AGN Eddington ratio $L/L_{Edd} \propto L/ M_{BH}$,  assuming that the X--ray luminosity represents the nuclear luminosity and the bulge luminosity is proportional to the bulge mass (and therefore to  M$_{BH}$).\\
Hence, we can guess a correlation between X/O ratio and the energy production efficiency
of the AGN measured by the Eddington ratio. Under these assumptions, the galaxy light concentration vs.
X/O flux ratio relation traces the correlation between the nuclear BH mass and the Eddington ratio. 
This result could therefore suggest that more concentrated or bulge-dominated (early-type) galaxies, having poor matter 
supply to feed the activity, have lower Eddington rates than those of late-type, with larger reservoires of the material for AGN feeding. \\

This suggested approach is consistent with the results obtained by \cite{ballo07}. They found that AGNs with large SMBH (M$_{BH}$\,$>$\,3\,$\times$\,10$^{6}$\,M$\odot$), have low X--ray luminosity and Eddington rate $\ll$ 1, and conversely, that smaller SMBH have higher luminosity in X--rays and Eddington rate $\approx$ 1, which corresponds to a more efficient accretion rate. \\
Our approach is also consistent with the results  of \cite{wu04}, which studied the
M$_{BH}$ and Eddington rates of a sample of 135 double--peaked broad line AGNs. 
They obtained that if the separation between the line peaks 
decrease (this separation is directly related with the FWHM of 
the line, and correlated with the M$_{BH}$),  
the Eddington rate increases.\\
\cite{kawakatu07} found an anticorrelation between the mass of a SMBH and the infrared-to-Eddington luminosity 
ratio, L$_{IR}$/L$_{Edd}$, for a sample of type 1 ULIRGs and nearby QSOs, which again could be consistent with our 
approach. The anticorrelation is interpreted as a link between the mass of a SMBH and the rate of mass 
accretion onto a SMBH, normalized by the AGN Eddington rate, which indicates that the growth of the BH is 
determinated by the external mass supply process, and not the AGN Eddington-limited mechanism, changing its mass 
accretion rate from super-Eddington to sub-Eddington.  \\
Nevertheless, as pointed out in Section \ref{virgo_millennium} there are observational examples of outliers of this anticorrelation, showing that massive elliptical galaxies can have high accreting BHs, and that late-type galaxies can host weakly accreting engines \citep[e.g.][]{churazov05,desroches09}.

\section{Acknowledgements}

We thank the anonymous referees for a number of valuable comments and suggestions which contributed to improve the quality of this paper. \\
This work was supported by the Spanish {\it Plan Nacional de
Astronom\'\i a y Astrof\'\i sica} under grants AYA2008-06311-C02-01 and AYA2008-06311-C02-02.
JIGS acknowledges financial support from the Spanish Ministerio 
de Educaci\'on y Ciencia under grant AYA2005-00055.\\
We acknowledge support from the Faculty of the European Space Astronomy Centre (ESAC).\\
We thank to SXDS team for making available their optical data to the astronomical community.\\
The Millennium II Simulation databases used in this paper and the web application providing
online access to them were constructed as part of the activities of the German Astrophysical Virtual
Observatory (GAVO). We thank to GAVO and Millennium Simulation team for making public their databases.\\


\bibliographystyle{aa}

\end{document}